\documentclass{article}
\usepackage{spconf,amsmath,graphicx,hyperref}
\usepackage{booktabs,multirow} 
\usepackage{enumitem}
\usepackage{amssymb}
\usepackage{amsfonts}
\usepackage{url}
\usepackage{graphicx}

\title{Behind the Scenes: Mechanistic Interpretability of LoRA-adapted Whisper for Speech Emotion Recognition}

\name{Yujian Ma $^{1}$ \qquad Xikun Lu$^{1}$ \qquad Jinqiu Sang$^{2*}\thanks{*Corresponding authors:  Jinqiu Sang (jqsang@mail.ecnu.edu.cn); Ruizhe Li (ruizhe.li@abdn.ac.uk).}$ \qquad Xianquan Jiang$^{3}$\qquad Ruizhe Li$^{4*}$
}
  
\address{$^1$ Shanghai Institute of Artificial Intelligence for Education, East China Normal University, China \\
$ ^2$School of Computer Science and Technology, East China Normal University,  China\\$^3$Boin Hearing Technology (Shanghai) Co., LTD, China\\$^4$Department of Computing Science, University of Aberdeen, UK}
\begin{document}
\ninept
\maketitle
\begin{abstract}
Large pre-trained speech models such as Whisper offer strong generalization but pose significant challenges for resource-efficient adaptation. Low-Rank Adaptation (LoRA) has become a popular parameter-efficient fine-tuning method, yet its underlying mechanisms in speech tasks remain poorly understood. In this work, we conduct the first systematic mechanistic interpretability study of LoRA within the Whisper encoder for speech emotion recognition (SER). Using a suite of analytical tools, including layer contribution probing, logit-lens inspection, and representational similarity via singular value decomposition (SVD) and centered kernel alignment (CKA), we reveal two key mechanisms: a \textit{delayed specialization} process that preserves general features in early layers before consolidating task-specific information, and a \textit{forward alignment, backward differentiation} dynamic between LoRA's matrices. Our findings clarify how LoRA reshapes encoder hierarchies, providing both empirical insights and a deeper mechanistic understanding to design efficient and interpretable adaptation strategies in large speech models. 
\end{abstract}
\section{Introduction}
\label{sec:intro}
Large-scale pre-trained models have fundamentally transformed speech and language processing. Models such as Whisper~\cite{whisper}, Wav2Vec2~\cite{wav2vec}, and HuBERT~\cite{hubert} provide general-purpose representations that are widely adapted to diverse tasks, from automatic speech recognition (ASR) to SER. Yet this versatility comes at a cost: the heavy computational and storage demands of full fine-tuning often hinder their deployment in resource-constrained settings. This tension has motivated the development of Parameter-Efficient Fine-Tuning (PEFT) methods~\cite{peft_survey,peft-ada}, which aim to adapt large encoders with minimal additional parameters.

Among PEFT approaches, LoRA~\cite{lora} has demonstrated strong effectiveness across multiple modalities. By injecting trainable low-rank matrices into frozen backbones, LoRA drastically reduces training costs while preserving the representational power of pre-trained models. Despite this practical success, most previous studies have focused on reporting performance improvements, while the underlying mechanisms behind LoRA’s effectiveness remain underexplored. Addressing this interpretability gap is critical for moving beyond empirical gains toward principled design of efficient fine-tuning strategies.

SER is not only a practically important task in affective computing but also a compelling domain for investigating parameter-efficient adaptation. Unlike ASR, which relies primarily on lexical information, SER is a classification task that depends on prosodic and paralinguistic cues reflecting high-level affective states~\cite{ser_survey}, which makes it well suited to analyze how LoRA's low-rank adaptation reshapes discriminative representations. The successful adaptation of Whisper~\cite{whisper_finetune} and Wav2Vec2~\cite{wav2vec_finetune} to SER further underscores the dual importance of enhancing performance and understanding underlying mechanisms in affective speech modeling.

Despite the growing popularity of parameter-efficient tuning, the internal mechanisms by which LoRA reshapes large speech encoders remain poorly understood, particularly for classification-oriented tasks such as SER. To bridge this gap, we present a systematic mechanistic study of LoRA within the Whisper encoder, leveraging probing and representational analyses to examine its adaptation process. Our code is publicly available\footnote{https://github.com/harryporry77/Behind-the-Scenes}.

\section{Related Work}
\label{sec:related}
\subsection{Mechanistic Interpretability}
Mechanistic interpretability (MI) is an emerging field that seeks to reverse-engineer the internal computations of deep networks to understand the algorithms they have learned~\cite{mi_survey}. In the domain of natural language processing (NLP), MI studies have shed light on how large language models (LLMs) encode complex behaviors such as arithmetic and reasoning~\cite{li2024anchored,li2025attributing}. For instance, Csordás et al. investigated the efficient use of network depth, revealing that in deep LLMs, the latter half of the network is often underutilized, primarily refining the output probability distribution rather than performing novel computations~\cite{layer_depth}. Building on this analytical framework, the `Backward Lens' study extended it to the backward pass, revealing a two-phase 'imprint and shift' mechanism for how models store new knowledge in their feed-forward layers~\cite{backward_lens}. Recent multimodal efforts, such as AudioLens~\cite{audiolens}, have further applied these principles to understand auditory perception in large audio-language models. However, within the speech processing community, MI remains limited, with most work focusing on probing intermediate representations rather than the intricate adaptation mechanisms of fine-tuning strategies like LoRA.

\subsection{Speech Emotion Recognition}
SER has evolved significantly, progressing from systems based on hand-crafted features~\cite{hand_features} to those leveraging powerful self-supervised encoders like Wav2Vec2~\cite{wav2vec} and HuBERT~\cite{hubert}, and more recently, large pre-trained models such as Whisper~\cite{whisper}. This evolution has led to notable improvements in cross-speaker and cross-domain robustness~\cite{ser_survey,hu2023mir}. Task-specific fine-tuning approaches, including those that incorporate metadata-enhanced strategies~\cite{downstream} and domain generalization for Whisper~\cite{whisper_finetune}, have shown promising results. Nevertheless, the substantial computational and storage costs associated with full fine-tuning necessitate the exploration of more resource-efficient alternatives~\cite{zeng2021affective,peft_ser}.

\subsection{Parameter-Efficient Fine-Tuning}
PEFT methods~\cite{peft_survey,peft-ada} offer a compelling solution by adapting frozen backbones with a minimal number of trainable parameters. Popular approaches include adapters, prefix-tuning, and prompt-tuning, among which LoRA~\cite{lora} has demonstrated exceptional effectiveness. While adapters may introduce inference latency and prefix-tuning can be unstable for speech tasks, LoRA achieves remarkable efficiency without sacrificing performance by injecting low-rank updates. Its efficacy has been confirmed in various applications, including Whisper for ASR~\cite{lora-whisper}, yet the precise inner adaptation dynamics of LoRA for complex tasks like SER remain largely unexplored.

\subsection{Research Gap}
In summary, existing literature has made significant strides in three distinct areas: mechanistic interpretability in NLP, the application of large pretrained encoders for SER, and the development of PEFT methods for speech. However, the intersection of these fields, specifically the mechanistic interpretability of LoRA’s adaptation in speech emotion recognition, has not been systematically investigated. This work is dedicated to bridging this critical research gap by analyzing how LoRA reshapes the representational hierarchies and optimization dynamics within the Whisper for the SER.

\section{Experiment}
\label{sec:experiment}
SER is formulated as a 4-class task (\textit{anger}, \textit{happiness}, \textit{neutrality}, \textit{sadness}) with WhisperForAudioClassification as the backbone. We conduct experiments on the IEMOCAP~\cite{iemocap} dataset using a standard speaker-independent 10-fold cross-validation. Unweighted recall (UAR) and weighted recall (WAR) serve as our evaluation metrics; UAR provides a robust measure for imbalanced datasets, while WAR reflects overall accuracy. For our mechanistic analysis, we use the Whisper-large-v2 encoder and the NNsight~\cite{nnsight} library, with analysis based on a hierarchical sample of 100 examples from a validation set (25 from each emotion class). We freeze the entire Whisper encoder (including LayerNorm and positional embeddings). For LoRA, only $\{A,B\}$ and the classification head are trainable; for the Frozen-Encoder baseline, only the same head is trained.

\subsection{LoRA Fine-tuning}
\label{sec:lora_result}
We adopt LoRA as our primary fine-tuning method, which approximates weight updates via a low-rank decomposition, $\Delta W = BA$. Only the low-rank matrices $A \in \mathbb{R}^{r \times d}$ and $B \in \mathbb{R}^{d \times r}$ are trainable, significantly reducing the number of parameters. We attach LoRA modules to the attention projections, a standard and effective strategy. Unless otherwise noted, we use a configuration of r=32 and apply a dropout of 0.1, with a trainable classification head. We use AdamW with a fixed training budget and report mean$\pm$std over the 10 folds. Hyperparameters are provided in released code.

\begin{table}[t]
\centering
\caption{SER performance on IEMOCAP under speaker-independent 10-fold cross-validation. Results are mean $\pm$ std.}
\label{tab:iemocap_results}
\footnotesize
\setlength{\tabcolsep}{2pt}
\begin{tabular}{l c c c c}
\toprule
\multirow{2}{*}{Model} & \multicolumn{2}{c}{LoRA} & \multicolumn{2}{c}{Frozen-Encoder} \\
\cmidrule(lr){2-3} \cmidrule(lr){4-5}
 & UAR & WAR & UAR & WAR \\
\midrule
tiny & $0.670\pm0.026$ & $0.664\pm0.028$ & $0.485\pm0.033$ & $0.502\pm0.028$ \\
base & $0.702\pm0.025$ & $0.692\pm0.025$ & $0.517\pm0.036$ & $0.530\pm0.029$ \\
small & $0.728\pm0.034$ & $0.723\pm0.036$ & $0.545\pm0.036$ & $0.558\pm0.036$ \\
medium & $0.758\pm0.030$ & $0.756\pm0.031$ & $0.638\pm0.037$ & $0.641\pm0.032$ \\
large-v2 & $\mathbf{0.774\pm0.026}$ & $\mathbf{0.768\pm0.035}$ & $0.582\pm0.044$ & $0.588\pm0.041$ \\
large-v3 & $0.767\pm0.034$ & $0.763\pm0.036$ & $0.433\pm0.031$ & $0.459\pm0.036$ \\
\bottomrule
\end{tabular}
\end{table}

As shown in Table~\ref{tab:iemocap_results}, LoRA consistently and significantly outperforms the frozen-encoder baselines across all model sizes. The Whisper-large-v2 with LoRA achieves the best performance, with a UAR of 0.774 and WAR of 0.768, demonstrating substantial improvements over its frozen counterpart. The performance of LoRA scales consistently with model size, highlighting its ability to effectively leverage larger pre-trained representations. In stark contrast, the frozen-encoder results show irregular patterns, which may reflect a fundamental incompatibility between the model's original ASR representations and the demands of the SER task. Our subsequent mechanistic analysis aims to clarify how LoRA effectively resolves this representational conflict.

\subsection{Layer Contribution Probing}
\label{subsec:layer_contribution}

\begin{figure}[t]
    \centering
    \includegraphics[width=\linewidth]{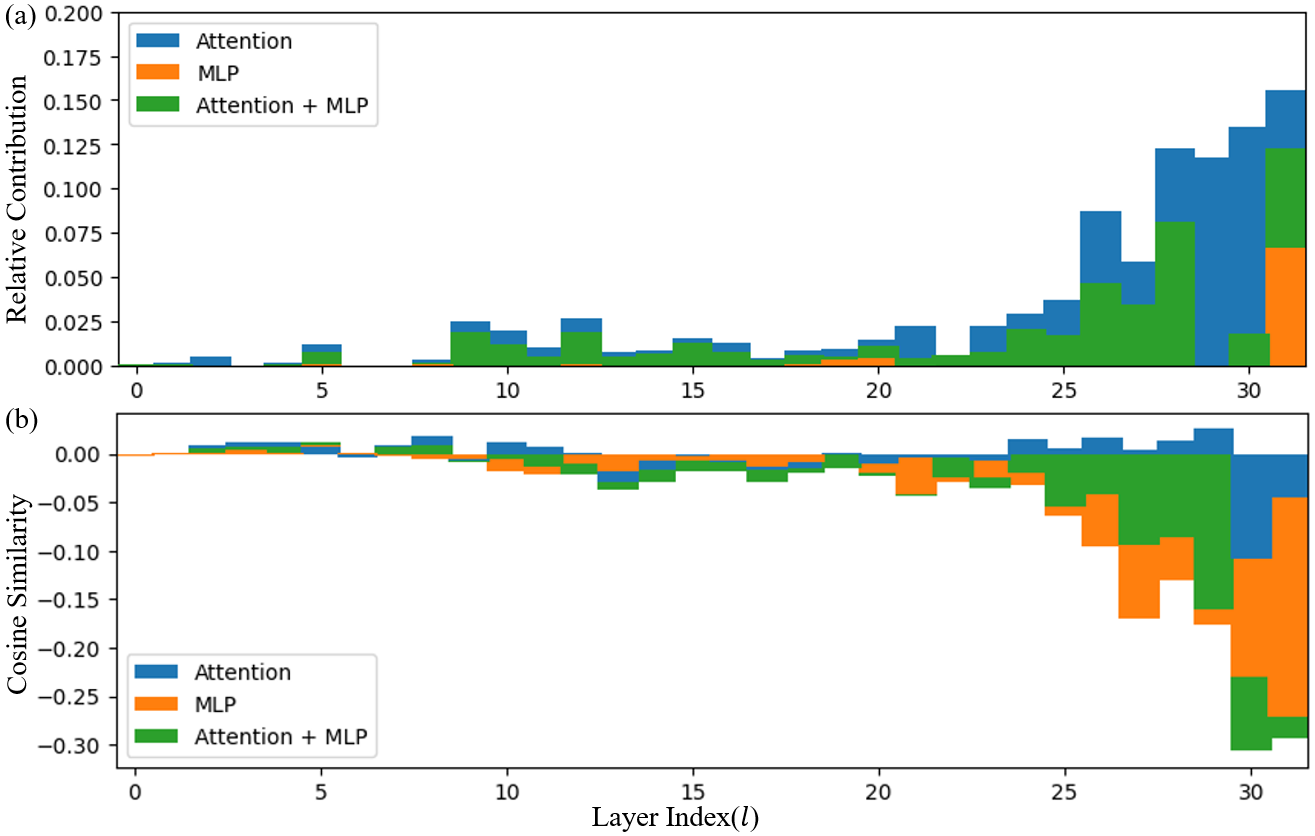}
    \caption{Layer-wise differences (LoRA minus frozen) on the encoder residual stream: (a) mean relative contribution and (b) cosine similarity for self-attention, MLP, and their sum.}
    \label{fig:layer_contribution}
\end{figure}

Our analysis of layer-wise contributions is conceptually inspired by recent work on analyzing the depth efficiency of language models~\cite{layer_depth}. We extend this methodology to the LoRA-adapted Whisper encoder by decomposing each Transformer block into its self-attention and MLP sublayers and measuring their effects on the residual stream. For a given layer $\ell$ with residual state $\mathbf{h}_\ell$, attention output $\mathbf{a}_\ell$, and MLP output $\mathbf{m}_\ell$, we compute the relative contribution ratios $\|\mathbf{a}_\ell\|_2 / \|\mathbf{h}_\ell\|_2$ and $\|\mathbf{m}_\ell\|_2 / \|\mathbf{h}_\ell\|_2$. We also measure the directional alignment of these outputs with the residual stream using cosine similarity: $\cos(\mathbf{a}_\ell,\mathbf{h}_\ell)$ and $\cos(\mathbf{m}_\ell,\mathbf{h}_\ell)$. To isolate LoRA’s specific effect, we report the per-layer differences $\Delta=(LoRA-frozen)$ averaged over the evaluation data, as shown in Fig.~\ref{fig:layer_contribution}.

While LoRA introduces negligible changes in the early layers, its relative contribution increases substantially towards the top of the encoder, indicating a depth-specific adaptation strategy. In these deeper layers, the attention sublayer’s contribution grows more than the MLP's, suggesting a primary role for LoRA in sharpening temporal focusing and long-range integration. This effect is further complemented by the combined Attention+MLP curve in the top layers, which evidences a synergistic interplay by exceeding the contribution of either component alone. The increasingly negative cosine similarity at higher layers indicates that the updates introduced by LoRA, which are directly applied to the Attention layers, are progressively passed to the subsequent layers. This suggests that the model's self-attention mechanism, having been finely tuned by LoRA to focus on emotional features, actively introduces signals that are counter-directional to the residual stream's information flow in the MLP layers. This is a critical mechanism for effective fine-tuning: by introducing these ``subtractive" or ``corrective" signals, LoRA is able to suppress or filter out features from the frozen backbone that are irrelevant or distracting to the new SER task. This targeted suppression allows the model to re-allocate its representational capacity, thereby emphasizing and reinforcing the task-relevant emotional features for a more robust and decisive final prediction.

\subsection{Logit-Lens Inspection}
\label{sec:logitlen_analysis}
To understand how discriminative information propagates through the encoder hierarchy, we conduct a Logit-Lens analysis~\cite{logit_lens}. The core idea of this method is to inspect the "mind" of the model by projecting the intermediate representations of each layer directly to the final output space. This allows us to observe the layer-wise evolution of the model's predictions. For each layer $\ell$, we extract the representation ${\mathbf{h}}_\ell$, then apply the trained projector and classifier to obtain intermediate logits $\mathbf{z}_\ell = \text{Classifier}(\text{Project}({\mathbf{h}}_\ell))$.

We quantify the alignment between intermediate and final predictions using two complementary metrics:
\begin{enumerate}[leftmargin=*]
    \item \textbf{KL Divergence}: $D_{KL}(\pi_\ell \| \pi_L) = \sum_c \pi_\ell(c) \log \frac{\pi_\ell(c)}{\pi_L(c)}$, where $\pi_\ell = \text{softmax}(\mathbf{z}_\ell)$ and $\pi_L$ is the final prediction. Lower values indicate better alignment.
    \item \textbf{Prediction Overlap}: $O_\ell = \mathbb{I}[\arg\max(\mathbf{z}_\ell) = \arg\max(\mathbf{z}_L)]$, measuring whether intermediate and final predictions agree. Higher values indicate stronger consistency.
\end{enumerate}

\begin{figure}[t]
\centering
\includegraphics[width=1.0\linewidth]{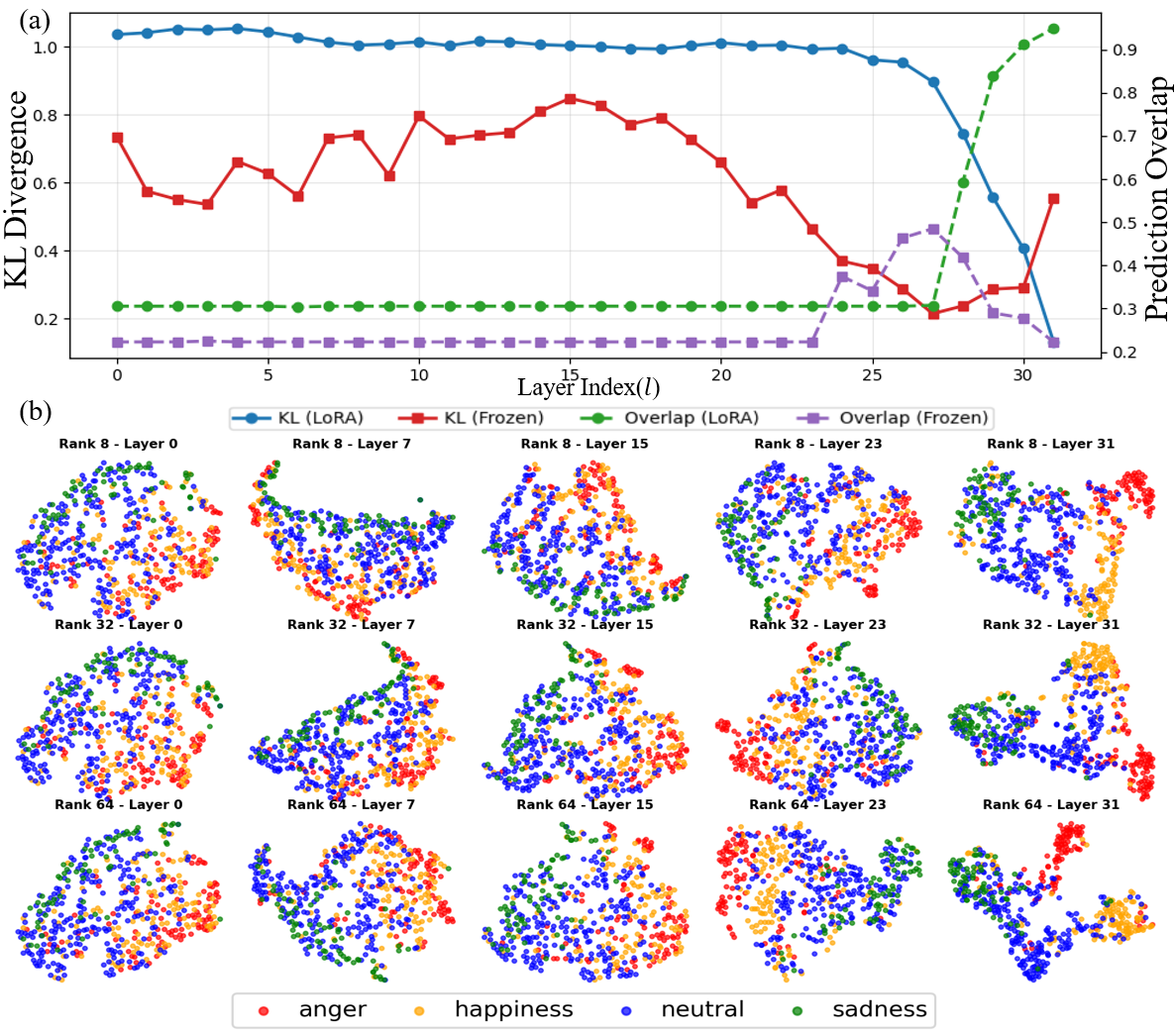}
\caption{Layer-wise analysis of LoRA's internal representations. (a) Logit-Lens analysis. (b) t-SNE visualization across different LoRA ranks.}
\label{fig:logit_len_tsne}
\end{figure}

As shown in Fig.~\ref{fig:logit_len_tsne}a, the two models exhibit distinct depth-wise alignment patterns. The frozen encoder shows volatile KL curves in early layers, drops to a minimum near layer 27, then rebounds at the top. This suggests that while Whisper possesses latent emotion-related capacity, achieving strongest emotional expression at the KL minimum, the original ASR objective creates inherent instability requiring downstream classifier compensation.

In contrast, the LoRA-adapted encoder employs a "delayed decision-making" strategy that maintains a relatively flat and high KL across the early and middle layers, then undergoes a pronounced late-stage drop converging to a very low value at the top. Rather than producing the chaotic signals observed in the frozen encoder, LoRA preserves a "stable yet unspecialized" state in the early layers, laying a smooth foundation for final specialization. This strategy allows LoRA to avoid premature commitment to task-specific representations while ensuring that the eventual specialization is both decisive and robust. This observation corroborates our findings in Section~\ref{subsec:layer_contribution}, as both the Logit-Lens and Layer Contribution analyses converge on a shared conclusion: LoRA's most significant contributions are concentrated in the deeper layers of the encoder.

The prediction-overlap curves further corroborate this delayed specialization mechanism. The frozen encoder begins to increase overlap earlier (around layer 23) but subsequently declines, mirroring the instability seen in its KL trajectory. The LoRA-adapted encoder maintains a higher and steady overlap up to the mid–upper layers, then shows a sharp rise starting around layer 27, quickly approaching near-perfect agreement at the top. This delayed but decisive top-1 prediction consolidation demonstrates LoRA's ability to defer critical decisions until the optimal moment, when sufficient task-relevant information has been accumulated and refined.

Taken together, these trends reveal that LoRA fundamentally restructures the information flow within the encoder: rather than relying on the original model's unstable emotional representations, it establishes a controlled, late-stage specialization process that preserves general representations in earlier layers while performing targeted, robust specialization where it is most effective. This controlled, late-stage specialization process not only improves performance but also enhances the stability and interpretability of the model's decision-making process.

\subsection{Rank-Based Representation}
\label{sec:tsne_analysis}

Our t-SNE analysis across different LoRA ranks ($r=8, 32, 64$) reveals a systematic improvement in emotion clustering quality with increasing rank capacity. As shown in Fig.~~\ref{fig:logit_len_tsne}b, deeper layers with $r=64$ achieve the most distinct and well-separated emotion boundaries, while maintaining a consistent depth-wise evolution from mixed to specialized representations. This pattern reinforces our delayed decision-making hypothesis: while the fundamental specialization mechanism remains consistent across ranks, LoRA's representational capacity directly determines the final clustering quality and emotion separability. Low-rank configurations ($r=8$) demonstrate a fundamental limitation, exhibiting persistent inter-class overlap even in the deepest layers, suggesting minimal rank capacity constrains the model's ability to achieve complete emotion separation.

An emotion-specific analysis reveals a hierarchy of rank-sensitivity. Neutral emotions maintain stable clustering across all ranks, likely due to their high frequency in the original Whisper training data. Sadness shows moderate rank-sensitivity, forming coherent clusters at $r=32$ and above, but fragmenting at $r=8$. In contrast, anger and happiness require higher representational capacity ($r=64$) for clear separation, with $r=8$ configurations showing severe confusion. This hierarchy—neutral $<$ sadness $<$ anger $<$ happiness—suggests that the recognition of positive emotions demands the highest level of LoRA sophistication, likely due to their nuanced acoustic expressions and the need for fine-grained feature discrimination.


\subsection{Singular Value Decomposition Analysis}
\label{sec:svd_analysis}
\begin{figure}[t]
\centering
\includegraphics[width=1.02\linewidth]{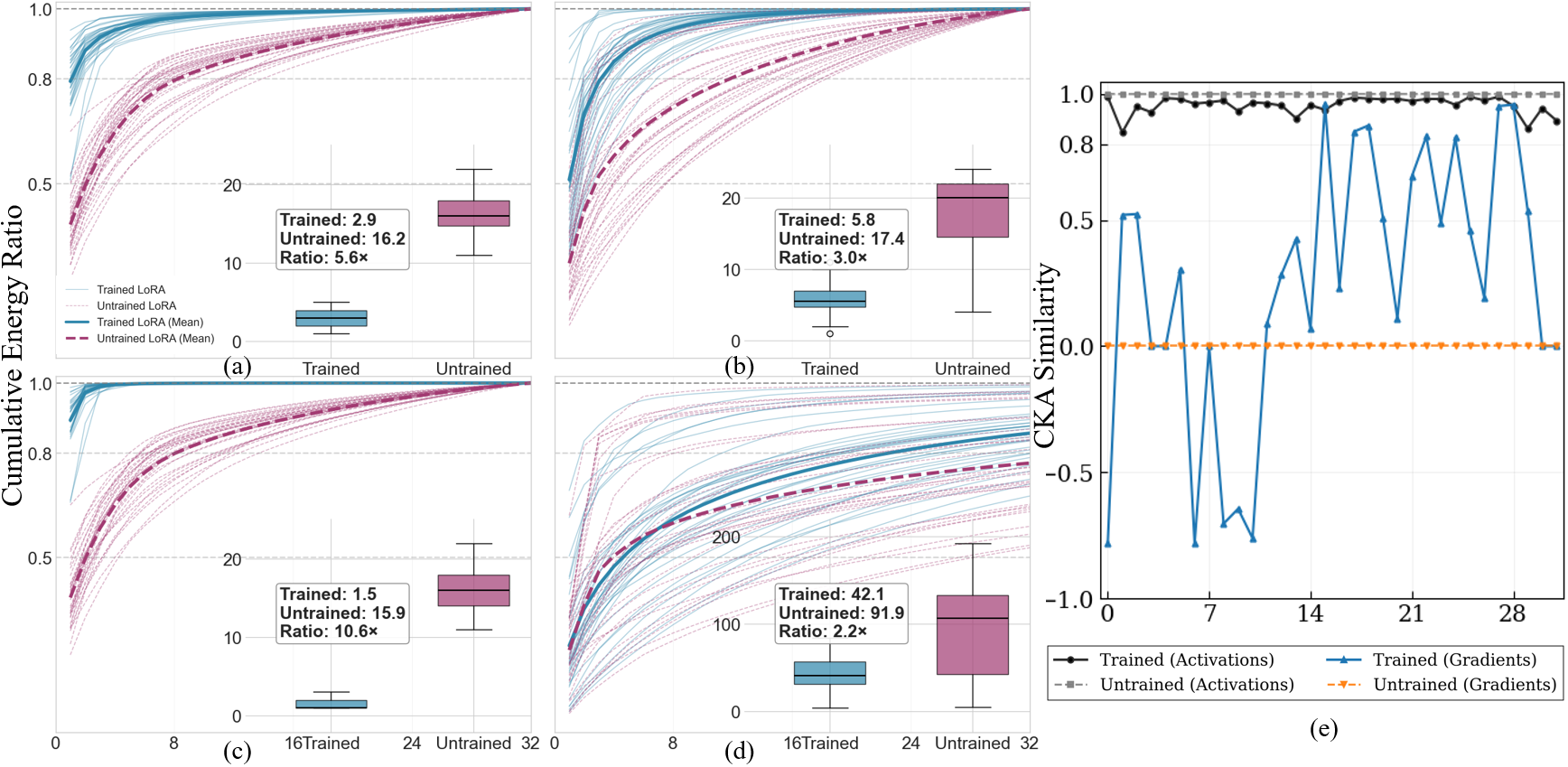}
\caption{Analysis of LoRA's internal dynamics. (a--d) SVD-based effective-rank curves for LoRA$_A$/LoRA$_B$ activations and their gradients, comparing trained LoRA (solid) with random initialization (dashed). The y-axis is cumulative energy ratio and the x-axis is the number of singular components. (e) CKA between LoRA$_A$ and LoRA$_B$ for activations and gradients.}
\label{fig:svd_cka}
\end{figure}
To investigate the intrinsic dimensionality and learning dynamics of LoRA, we perform SVD on the activation matrices produced by LoRA$_A$ and LoRA$_B$, and on their corresponding gradient matrices, across all layers. The ``LoRA (mean)'' is computed by averaging the spectra across layers (and attention heads when applicable), providing a global perspective on learning dynamics. Throughout, we use an energy-based effective-rank view: we report how many singular components are required to reach a fixed cumulative energy threshold (90\%), and the reported values can be non-integer due to averaging across layers. By comparing a trained LoRA model against a randomly initialized counterpart, we analyze how each component concentrates energy within its dominant singular components. This analysis suggests complementary roles between LoRA's components for compression and reconstruction.

LoRA$_A$ acts as a compression branch. Compared with random initialization, its trained activation spectrum (Fig.~\ref{fig:svd_cka}a) reaches the 90\% cumulative-energy threshold with only 2.9 singular components on average (5.5$\times$ more concentrated), and its gradients (Fig.~\ref{fig:svd_cka}b) also become more concentrated (1.7$\times$), suggesting optimization focuses on a small set of dominant directions. In contrast, LoRA$_B$ acts as a reconstruction branch. Compared with random initialization, its activations (Fig.~\ref{fig:svd_cka}c) become strongly concentrated after training (10.4$\times$), while its gradients (Fig.~\ref{fig:svd_cka}d) require more components to reach the same threshold, indicating more dispersed gradient energy and suggesting more diverse update directions. Notably, dispersed gradient energy does not imply a higher algebraic rank of $\Delta W$, since updates remain constrained by the low-rank parameterization.

This complementary learning behavior, where LoRA$_A$ concentrates energy into a low-dimensional subspace while LoRA$_B$ preserves richer gradient diversity, helps explain LoRA's parameter efficiency. Overall, the results are consistent with the adaptation being effectively low-dimensional under this task setting.

\subsection{Centered Kernel Alignment Analysis}
\label{sec:cka_analysis}

To quantify representational similarity between LoRA components, we employ CKA~\cite{CKA}, a normalized alignment measure for comparing neural representations across layers.  It is robust to isotropic scaling and orthogonal transformations of the representations, and is therefore more appropriate than raw cosine similarity for high-dimensional feature comparisons.  We compute  linear CKA layerwise for both forward activations and backward gradient signals.

\begin{equation}
\text{CKA}(\mathbf{X}, \mathbf{Y}) = \frac{\text{HSIC}(\mathbf{K}, \mathbf{L})}{\sqrt{\text{HSIC}(\mathbf{K}, \mathbf{K}) \cdot \text{HSIC}(\mathbf{L}, \mathbf{L})}}
\end{equation}

where $\mathbf{X}\in\mathbb{R}^{n\times d_x}$ and $\mathbf{Y}\in\mathbb{R}^{n\times d_y}$ are representation matrices with rows as examples and columns as features. $\mathbf{K}$ and $\mathbf{L}$ are centered Gram matrices derived from $\mathbf{X}$ and $\mathbf{Y}$, and HSIC~\cite{HSIC} is a kernel-based dependence measure computed from these centered Gram matrices. For forward activations, $\mathbf{X}$ and $\mathbf{Y}$ correspond to LoRA$_A$/LoRA$_B$ output activations at the same layer. For backward signals, $\mathbf{X}$ and $\mathbf{Y}$ correspond to the gradients of the loss with respect to these LoRA output activations.

As depicted in Fig.~\ref{fig:svd_cka}e, LoRA exhibits a clear forward and backward asymmetry. Forward activations of LoRA$_A$ and LoRA$_B$ remain highly aligned (0.8--1.0) across layers, indicating that the two components induce very similar representational structure during the forward pass. In contrast, gradient alignment is strongly layer-dependent and varies substantially, suggesting that the two components can receive differentiated optimization signals even when their forward representations are similar. 

SVD and CKA are complementary: the former characterizes energy concentration within each component, and the latter quantifies layerwise alignment between components. Together they indicate compression by LoRA$_A$ and reconstruction by LoRA$_B$ with forward alignment and backward differentiated dynamics.

\section{CONCLUSION}
\label{sec:conclusion}
This paper provides the first systematic mechanistic analysis of LoRA's role within the Whisper encoder for SER. Our findings reveal two key mechanisms that contribute to LoRA's superior performance: a delayed specialization strategy, which maintains stable representations in early layers to enable decisive late-stage consolidation, and a forward alignment, backward differentiation dynamic between LoRA's matrices. These insights not only elucidate LoRA's effectiveness but also offer a deeper understanding of how low-rank fine-tuning methods fundamentally restructure the information flow within deep speech models. Our work establishes a foundation for designing more efficient and interpretable adaptation strategies across various speech tasks, and we believe our analysis framework can serve as a blueprint for investigating whether these dynamics generalize to other models and datasets.

\vfill\pagebreak

\section{ACKNOWLEDGEMENTS}
This work was supported by the National Natural Science Foundation of China (Grant No. 12411530075), the Royal Society (Grant No. IEC\textbackslash NSFC\textbackslash 233558) and Shanghai Municipal Commission of Economy and Informatization (No. 2024-GZL-RGZN-02001).

\bibliographystyle{IEEEbib}
\bibliography{strings,refs}

\begin{thebibliography}{10}

\bibitem{whisper}
Alec Radford, Jong~Wook Kim, Tao Xu, Greg Brockman, Christine McLeavey, and Ilya Sutskever,
\newblock ``Robust speech recognition via large-scale weak supervision,''
\newblock in {\em ICML}. PMLR, 2023, pp. 28492--28518.

\bibitem{wav2vec}
Alexei Baevski, Yuhao Zhou, Abdelrahman Mohamed, and Michael Auli,
\newblock ``wav2vec 2.0: A framework for self-supervised learning of speech representations,''
\newblock {\em NeurIPS}, vol. 33, pp. 12449--12460, 2020.

\bibitem{hubert}
Wei-Ning Hsu, Benjamin Bolte, Yao-Hung~Hubert Tsai, Kushal Lakhotia, Ruslan Salakhutdinov, and Abdelrahman Mohamed,
\newblock ``Hubert: Self-supervised speech representation learning by masked prediction of hidden units,''
\newblock {\em IEEE/ACM transactions on audio, speech, and language processing}, vol. 29, pp. 3451--3460, 2021.

\bibitem{peft_survey}
Luping Wang, Sheng Chen, Linnan Jiang, Shu Pan, Runze Cai, Sen Yang, and Fei Yang,
\newblock ``Parameter-efficient fine-tuning in large language models: a survey of methodologies,''
\newblock {\em Artificial Intelligence Review}, vol. 58, no. 8, pp. 227, 2025.

\bibitem{peft-ada}
Nakamasa Inoue, Shinta Otake, Takumi Hirose, Masanari Ohi, and Rei Kawakami,
\newblock ``Elp-adapters: Parameter efficient adapter tuning for various speech processing tasks,''
\newblock {\em IEEE/ACM Transactions on Audio, Speech, and Language Processing}, 2024.

\bibitem{lora}
Edward~J Hu, Yelong Shen, Phillip Wallis, Zeyuan Allen-Zhu, Yuanzhi Li, Shean Wang, Lu~Wang, Weizhu Chen, et~al.,
\newblock ``Lora: Low-rank adaptation of large language models.,''
\newblock {\em ICLR}, vol. 1, no. 2, pp. 3, 2022.

\bibitem{ser_survey}
Siddique Latif, Rajib Rana, Sara Khalifa, Raja Jurdak, Junaid Qadir, and Bj{\"o}rn Schuller,
\newblock ``Survey of deep representation learning for speech emotion recognition,''
\newblock {\em IEEE Transactions on Affective Computing}, vol. 14, no. 2, pp. 1634--1654, 2021.

\bibitem{whisper_finetune}
Erik Goron, Lena Asai, Elias Rut, and Martin Dinov,
\newblock ``Improving domain generalization in speech emotion recognition with whisper,''
\newblock in {\em ICASSP}. IEEE, 2024, pp. 11631--11635.

\bibitem{wav2vec_finetune}
Li-Wei Chen and Alexander Rudnicky,
\newblock ``Exploring wav2vec 2.0 fine tuning for improved speech emotion recognition,''
\newblock in {\em ICASSP}. IEEE, 2023, pp. 1--5.

\bibitem{mi_survey}
Daking Rai, Yilun Zhou, Shi Feng, Abulhair Saparov, and Ziyu Yao,
\newblock ``A practical review of mechanistic interpretability for transformer-based language models,''
\newblock {\em arXiv preprint arXiv:2407.02646}, 2024.

\bibitem{li2024anchored}
Ruizhe Li and Yanjun Gao,
\newblock ``Anchored answers: Unravelling positional bias in gpt-2's multiple-choice questions,''
\newblock {\em arXiv preprint arXiv:2405.03205}, 2024.

\bibitem{li2025attributing}
Ruizhe Li, Chen Chen, Yuchen Hu, Yanjun Gao, Xi~Wang, and Emine Yilmaz,
\newblock ``Attributing response to context: A jensen-shannon divergence driven mechanistic study of context attribution in retrieval-augmented generation,''
\newblock {\em arXiv preprint arXiv:2505.16415}, 2025.

\bibitem{layer_depth}
R{\'o}bert Csord{\'a}s, Christopher~D Manning, and Christopher Potts,
\newblock ``Do language models use their depth efficiently?,''
\newblock {\em arXiv preprint arXiv:2505.13898}, 2025.

\bibitem{backward_lens}
Shahar Katz, Yonatan Belinkov, Mor Geva, and Lior Wolf,
\newblock ``Backward lens: Projecting language model gradients into the vocabulary space,''
\newblock {\em arXiv preprint arXiv:2402.12865}, 2024.

\bibitem{audiolens}
Chih-Kai Yang, Neo Ho, Yi-Jyun Lee, and Hung-yi Lee,
\newblock ``Audiolens: A closer look at auditory attribute perception of large audio-language models,''
\newblock {\em arXiv preprint arXiv:2506.05140}, 2025.

\bibitem{hand_features}
Bj{\"o}rn Schuller, Gerhard Rigoll, and Manfred Lang,
\newblock ``Speech emotion recognition combining acoustic features and linguistic information in a hybrid support vector machine-belief network architecture,''
\newblock in {\em 2004 IEEE international conference on acoustics, speech, and signal processing}. IEEE, 2004, vol.~1, pp. I--577.

\bibitem{hu2023mir}
Yuchen Hu, Chen Chen, Ruizhe Li, Heqing Zou, and Eng~Siong Chng,
\newblock ``Mir-gan: Refining frame-level modality-invariant representations with adversarial network for audio-visual speech recognition,''
\newblock {\em arXiv preprint arXiv:2306.10567}, 2023.

\bibitem{downstream}
Zixiang Wan, Ziyue Qiu, Yiyang Liu, and Wei-Qiang Zhang,
\newblock ``Metadata-enhanced speech emotion recognition: Augmented residual integration and co-attention in two-stage fine-tuning,''
\newblock in {\em ICASSP}. IEEE, 2025, pp. 1--5.

\bibitem{zeng2021affective}
Chengkun Zeng, Guanyi Chen, Chenghua Lin, Ruizhe Li, and Zhi Chen,
\newblock ``Affective decoding for empathetic response generation,''
\newblock in {\em INLG}, 2021, pp. 331--340.

\bibitem{peft_ser}
Nineli Lashkarashvili, Wen Wu, Guangzhi Sun, and Philip~C Woodland,
\newblock ``Parameter efficient finetuning for speech emotion recognition and domain adaptation,''
\newblock in {\em ICASSP}. IEEE, 2024, pp. 10986--10990.

\bibitem{lora-whisper}
Zheshu Song, Jianheng Zhuo, Yifan Yang, Ziyang Ma, Shixiong Zhang, and Xie Chen,
\newblock ``Lora-whisper: Parameter-efficient and extensible multilingual asr,''
\newblock {\em arXiv preprint arXiv:2406.06619}, 2024.

\bibitem{iemocap}
Carlos Busso, Murtaza Bulut, Chi-Chun Lee, Abe Kazemzadeh, Emily Mower, Samuel Kim, Jeannette~N Chang, Sungbok Lee, and Shrikanth~S Narayanan,
\newblock ``Iemocap: Interactive emotional dyadic motion capture database,''
\newblock {\em Language resources and evaluation}, vol. 42, no. 4, pp. 335--359, 2008.

\bibitem{nnsight}
Jaden Fiotto-Kaufman, Alexander~R Loftus, Eric Todd, Jannik Brinkmann, Koyena Pal, Dmitrii Troitskii, Michael Ripa, Adam Belfki, Can Rager, Caden Juang, et~al.,
\newblock ``Nnsight and ndif: Democratizing access to open-weight foundation model internals,''
\newblock {\em arXiv preprint arXiv:2407.14561}, 2024.

\bibitem{logit_lens}
nostalgebraist,
\newblock ``Interpreting gpt: the logit lens,'' \url{https://www.lesswrong.com/posts/AcKRB8wDpdaN6v6ru/interpreting-gpt-the-logit-lens}, Aug. 2020.

\bibitem{CKA}
Simon Kornblith, Mohammad Norouzi, Honglak Lee, and Geoffrey Hinton,
\newblock ``Similarity of neural network representations revisited,''
\newblock in {\em ICML}. PMlR, 2019, pp. 3519--3529.

\bibitem{HSIC}
Arthur Gretton, Olivier Bousquet, Alex Smola, and Bernhard Sch{\"o}lkopf,
\newblock ``Measuring statistical dependence with hilbert-schmidt norms,''
\newblock in {\em International conference on algorithmic learning theory}. Springer, 2005, pp. 63--77.

\end{thebibliography}

\end{document}